\documentclass[prb,twocolumn,preprintnumbers,amsmath,amssymb,superscriptaddress,showpacs]{revtex4}
\usepackage{graphicx}
\usepackage{epsfig}
\usepackage{dcolumn} 
\usepackage{bm} 

\newcommand {\PrLSMO} {(La$_{0.4}$Pr$_{0.6}$)$_{1.2}$Sr$_{1.8}$Mn$_2$O$_7$}

\begin{document}

\title{Magneto-optical investigation of the field-induced spin-glass insulator to ferromagnetic metallic transition of the bilayer manganite \PrLSMO\ }

\author{J. Cao}
\email{cao@ion.chem.utk.edu} \affiliation{Department of Chemistry,
University of Tennessee, Knoxville, TN 37996}
\author{J. T. Haraldsen}
\affiliation{Department of Physics and Astronomy, University of
Tennessee, Knoxville, TN 37996}
\author{R. C. Rai}
\affiliation{Department of Chemistry, University of Tennessee,
Knoxville, TN 37996}
\author{S. Brown}
\affiliation{Department of Chemistry, University of Tennessee,
Knoxville, TN 37996}
\author{J. L. Musfeldt}
\affiliation{Department of Chemistry, University of Tennessee,
Knoxville, TN 37996}
\author{Y. J. Wang}
\affiliation{National High Magnetic Field Laboratory, Florida State
University, Tallahassee, FL 32310}
\author{X. Wei}
\affiliation{National High Magnetic Field Laboratory, Florida State
University, Tallahassee, FL 32310}
\author{M. Apostu}
\altaffiliation[Permanent address: ]{Al. I. Cuza University, Faculty
of Chemistry, Physical, Theoretical and Materials Chemistry, Carol
I, Iasi 700506 Romania}

\affiliation{Laboratoire de Physico-Chimie de l'Etat Solide, UMR
8648 Batiment 414, Universit$\acute e$ Paris-Sud. F-91405 Orsay,
France}

\author{R. Suryanarayanan}
\affiliation{Laboratoire de Physico-Chimie de l'Etat Solide, UMR
8648 Batiment 414, Universit$\acute e$ Paris-Sud. F-91405 Orsay,
France}
\author{A. Revcolevschi}
\affiliation{Laboratoire de Physico-Chimie de l'Etat Solide, UMR
8648 Batiment 414, Universit$\acute e$ Paris-Sud. F-91405 Orsay,
France}

\begin{abstract}

We measured the magneto-optical response of
(La$_{0.4}$Pr$_{0.6}$)$_{1.2}$Sr$_{1.8}$Mn$_2$O$_7$ in order to
investigate the microscopic aspects of the magnetic field driven
spin-glass insulator to ferromagnetic metal transition. Application
of a magnetic field recovers the ferromagnetic state with an overall
redshift of the electronic structure, growth of the bound carrier
localization associated with ferromagnetic domains, development of a
pseudogap, and softening of the Mn-O stretching and bending modes
that indicate a structural change. We discuss field- and
temperature-induced trends within the framework of the
Tomioka-Tokura global electronic phase diagram picture and suggest
that controlled disorder near a phase boundary can be used to tune
the magnetodielectric response. Remnants of the spin-glass insulator
to ferromagnetic metallic transition can also drive 300 K color
changes in (La$_{0.4}$Pr$_{0.6}$)$_{1.2}$Sr$_{1.8}$Mn$_2$O$_7$.

\end{abstract}


\maketitle \clearpage

\section{INTRODUCTION}

\begin{figure}
\includegraphics[width = 3.0in]{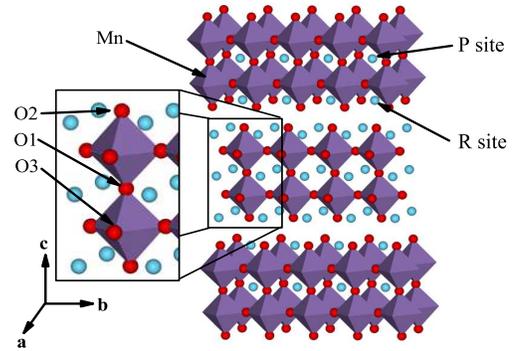}
\caption{\label{fig_LSMOstruc} (Color online) Crystal structure of
\PrLSMO\,, with O (red/black), Mn-containing octahedra
(purple/gray), and the rare/alkaline earth ions (light blue/light
gray). O occupies three different sites, and rare earth/alkaline
earth metal have different coordination numbers depending on whether
they occupy perovskite (P) or rock salt (R) sites. \cite{Kubota2000,
Apostu2001} Note that the unit cell along the $b$ axis has been
replicated to highlight the layered structure. }
\end{figure}

Substituted perovskite manganites have attracted considerable
attention due to their exotic magnetic, electronic, and optical
properties. These properties derive from the many competing ground
states of the complex phase diagram, strong coupling across
different energy scales, and the presence of an inhomogeneous
texture. \cite{Tokura2000, Dagotto2003, Chatterji2004} One
consequence of this complexity is that enormous physical property
changes can be induced by small chemical and physical perturbations.
The double-layer manganites of interest here derive from the
La$_{1.2}$Sr$_{1.8}$Mn$_2$O$_7$ parent compound, crystallizing in a
body-centered tetragonal structure (space group $I4/mmm$) as shown
in Fig.~\ref{fig_LSMOstruc}.\cite{Kubota2000, Apostu2001} These
bilayer manganites display a broad metallic regime, colossal
magnetoresistance, Jahn-Teller distortions, metal-insulator
transitions, and charge/orbital ordering.\cite{Moritomo1996,
Mitchell1997, Perring1997} These materials are therefore well-suited
for fundamental magnetodielectric properties
investigations.\cite{Cruz2005, Lorenz2004, Goto2004, Rogado2005,
Choi2004a, Choi2004b, Woodward2005, Rai2006, Lee2002,
Alexandrov2000, Jung2000a, Jung2000b, Okimoto1998, Okimoto1999,
Freitas2005} They are also useful to extend the Tomioka-Tokura
electronic phase diagram picture\cite{Tomioka2004} and test
oscillator strength sum rules. \cite{Alexandrov2000,Alexandrov1999}

Pr substitution of the La sites in La$_{1.2}$Sr$_{1.8}$Mn$_2$O$_7$
yields materials with chemical formula
(La$_{1-z}$Pr$_{z}$)$_{1.2}$Sr$_{1.8}$Mn$_2$O$_7$, providing an
opportunity to investigate the physical properties of bilayer
manganites as a function of disorder without changing the total hole
concentration.\cite{Apostu2001} Upon increasing Pr substitution
($z$=0, 0.2, and 0.4), the paramagnetic insulator (PMI) to
ferromagnetic metal (FMM) transition temperature, T$_c$, decreases
(120, 90, and 60 K, respectively).\cite{Matsukawa2003} At $z$=0.6,
the transition is quenched.\cite{Matsukawa2003} This material,
(La$_{0.4}$Pr$_{0.6}$)$_{1.2}$Sr$_{1.8}$Mn$_2$O$_7$, is the subject
of our present work. Pr substitution modifies the lattice constants
($c$/$a$ increases), causes a change in the Jahn-Teller distortion,
induces bound carrier localization in the far infrared, and modifies
the $e_g$ orbital occupancy.\cite{Matsukawa2003,
Wang2003,Woodward2004} Clearly, the degree of disorder  strongly
influences the physical properties. In fact, given that
(La$_{0.4}$Pr$_{0.6}$)$_{1.2}$Sr$_{1.8}$Mn$_2$O$_7$ is highly
disordered, the aforementioned low-temperature PMI state can likely
be considered to be a spin-glass insulator (SGI) as well, although
variable frequency ac susceptibility measurements are needed to
confirm this picture.\cite{Tomioka2004, Apostu2001}

The long-range ordered ferromagnetic state that is suppressed by
chemical pressure in
(La$_{0.4}$Pr$_{0.6}$)$_{1.2}$Sr$_{1.8}$Mn$_2$O$_7$ is recovered
under magnetic field. This recovery is evident in the H-T phase
diagram deduced from magnetization, magnetostriction, and
resistivity measurements. \cite{Gordon2001, Wagner2002,
Matsukawa2004, Matsukawa2005} Neutron scattering demonstrates that
the field-induced FMM state is very similar to the FMM state in the
double-layer parent compound,
La$_{1.2}$Sr$_{1.8}$Mn$_2$O$_7$.\cite{Moussa2004} Neutron
diffraction studies show that local structure (Mn-O bond distances
and Mn-O-Mn bond angles) and $e_g$  orbital occupancies
 change dramatically in  magnetic field, directly influencing
 electron hopping between Mn sites.\cite{Wang2003, Gukasov2005} Magneto-optical imaging suggests that
the high field FMM state is homogenous.\cite{Tokunaga2005}

In order to understand the interplay between spin, charge, lattice,
and orbital degrees of freedom, we investigated the magneto-optical
properties of (La$_{0.4}$Pr$_{0.6}$)$_{1.2}$Sr$_{1.8}$Mn$_2$O$_7$.
Application of a magnetic field recovers the ferromagnetic state
with an overall redshift of the electronic structure, growth of the
bound carrier localization associated with ferromagnetic domains,
and softening of the Mn-O stretching and bending modes. The high
field state is not, however, metallic in the conventional sense, and
the spectrum differs from that of the double-layer parent compound
La$_{1.2}$Sr$_{1.8}$Mn$_2$O$_7$ in that it displays a pseudogap. We
discuss field- and temperature-induced trends within the framework
of the Tomioka-Tokura global electronic phase diagram picture
\cite{Tomioka2004} and suggest that controlled disorder near a phase
boundary can be used to tune the magnetodielectric response. We also
employ these microscopic changes to extract the H-T phase diagram
and show that the low temperature lattice responds more slowly than
spin and charge.  Finally, we demonstrate that remnants of the
SGI-FMM transition
can drive 300 K color changes  in (La$_{0.4}$Pr$_{0.6}$)$_{1.2}$Sr$_{1.8}$Mn$_2$O$_7$.  

\section{Methods}

Single crystals of \PrLSMO\ were grown from sintered rods of same
nominal composition by the floating-zone technique, using a mirror
furnace.\cite{Apostu2001} Typical crystal dimensions were
$\approx$4$\times$5$\times$2 mm$^3$. They were cleaved to yield a
shiny surface corresponding to the $ab$ plane.

Near normal $ab$ plane  reflectance of \PrLSMO\ was measured over a
wide energy range (3.7 meV - 6.5 eV) using different spectrometers
including a Bruker 113V Fourier transform infrared spectrometer, a
Bruker Equinox 55 Fourier transform infrared spectrometer equipped
with an infrared microscope, and a Perkin Elmer Lambda 900 grating
spectrometer. The spectral resolution was 2 cm$^{-1}$ in the far and
middle-infrared and 2 nm in the near-infrared, visible, and
near-ultraviolet. Aluminum mirrors were used as references for all
measurements. Low temperature spectroscopies were carried out with a
continuous-flow helium cryostat and temperature controller. Optical
conductivity was calculated by a Kramers-Kronig analysis of the
measured reflectance.\cite{Wooten1972}

The magneto-optical properties of \PrLSMO\ were measured at the
National High Magnetic Field Laboratory (NHMFL) in Tallahassee, FL,
using a Bruker 113V Fourier transform infrared spectrometer equipped
with a 18 T superconducting magnet and a grating spectrometer
equipped with InGaAs and CCD detectors and a 33 T resistive magnet.
Experiments were performed at 4.2 K for $H\parallel c$. Selected
experiments were also carried out between 4.2 and 300 K in the
spectral range of 0.75 - 3 eV. Data were collected on both
increasing and decreasing magnetic field. Upsweep data were plotted
here, whereas hysteresis effects on optical properties are discussed
elsewhere.\cite{Choi2004a} After each field sweep, the samples were
heated to 80 K to erase the ``memory''. The field-induced changes in
the measured reflectance were studied by taking the ratio of
reflectance at each field and reflectance at zero field, i.e.,
[R($H$)/R($H$=0 T)]. To obtain the high field optical conductivity,
we renormalized the zero-field absolute reflectance with the
high-field reflectance ratios, and recalculated $\sigma$$_1$ using
Kramers-Kronig techniques.\cite{Wooten1972} Due to limited coverage
of the spectrometers at the NHMFL, the measured data were spliced
together with curve fitting techniques between $\sim$0.5 and 0.75
eV.

We employed standard color rendering techniques to visualize
temperature- and field-induced spectral
changes.\cite{Billmeyer2000,Musfeldt1993} Here, the absorption
coefficient data\cite{absorption} are ``matched" with the effective
absorption using a proportionality constant (which is typically on
the order of the pellet thickness times the loading).\cite{K} A
comparison of the absorption coefficient to the effective absorption
spectrum can be used to render color by integrating the product of
the spectrum with the well-known XYZ color matching functions to
determine the XYZ color values.\cite{Billmeyer2000} These XYZ values
are converted into RGB color values and then inverted to determine
the color of a material.\cite{Billmeyer2000} The final RGB values
allow color rendering.

\section{RESULTS AND DISCUSSION}

\subsection{Field Dependence of the Optical Spectra}

\begin{figure}
\includegraphics[width = 3.5 in]{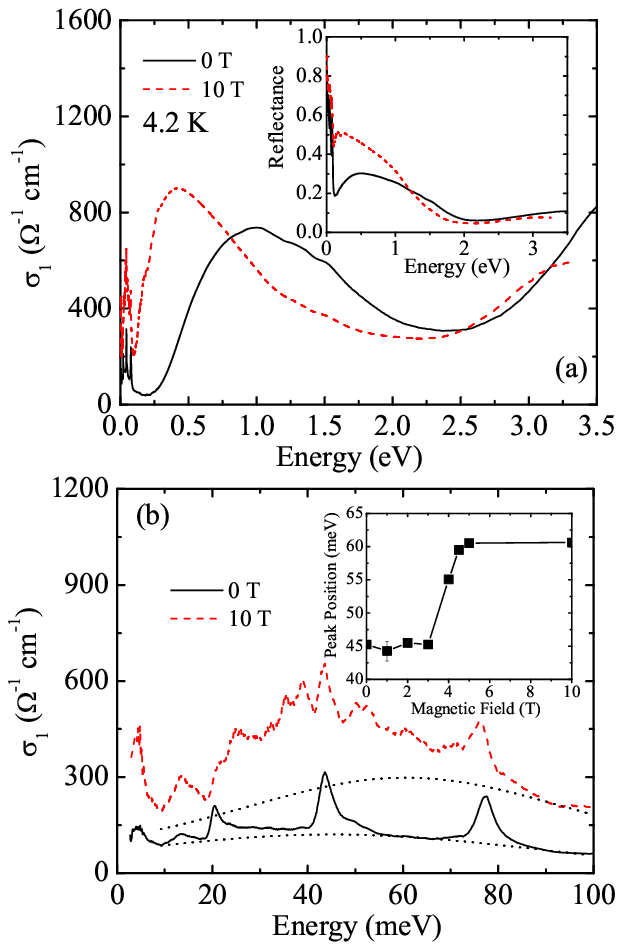}
\caption{\label{fig_PrCondReflPrFM} (Color online) (a) Optical
conductivity of \PrLSMO\ within the $ab$ plane under 0 (solid line)
and 10 T (dashed line) at 4.2 K ($H\parallel c$), extracted from a
Kramers-Kronig analysis of the measured reflectance data (inset).
(b) Close-up view of the 4.2 K optical conductivity within the $ab$
plane under 0 and 10 T magnetic fields ($H\parallel c$). Dotted
lines guide the eye to highlight the broad underlying bound carrier
localization. The inset shows the development of the bound carrier
localization (dotted lines), associated with ferromagnetic domains,
upon application of field. }
\end{figure}

Figure~\ref{fig_PrCondReflPrFM}(a) displays the $ab$ plane  optical
conductivity of \PrLSMO\ in the  low (SGI) and high field (FMM)
states at 4.2 K. The dominant effect of the applied field is to
redshift the oscillator strength, which is conserved within a few
percent. A similar field-induced redshift of the spectral weight has
been observed in other manganites.\cite{Jung2000a, Jung2000b,
Okimoto1998, Okimoto1999} In
(La$_{0.4}$Pr$_{0.6}$)$_{1.2}$Sr$_{1.8}$Mn$_2$O$_7$, the electronic
excitation near 1 eV in the SGI state arises from a combination of
Mn$^{3+}$ intra-atomic $d$ to $d$ excitations superimposed with
Mn$^{3+}$ to Mn$^{4+}$ ($d$ to $d$) inter-atomic charge transfer
excitations.\cite{Ishikawa1998,Lee2000,Woodward2004}
 This feature narrows and shifts
to 0.4 eV in the high field FMM state. With increasing magnetic
field, spectral weight also transfers to the lower energy region,
enlarging the bound carrier excitation in the far infrared. This
feature is a signature of ferromagnetic domain
formation\cite{Woodward2004} and is discussed in more detail below.
Note that the optical conductivity of \PrLSMO\ does not show typical
metallic behavior under any circumstances. Even at 10 T, where the
system is driven into the FMM state, a far-infrared bound carrier
excitation is observed rather than a Drude response. Manganites
generally display low dc
conductivities,\cite{Tokura2000,Chatterji2004,Dagotto2003} and
\PrLSMO\ is no exception. Transport measurements show that
$\sigma_{dc}$  is $\sim$10$^{-5}$ $\Omega^{-1}$ cm$^{-1}$ in the low
temperature insulating  state and $\sim$250 $\Omega^{-1}$ cm$^{-1}$
in the high field metallic state,\cite{Gordon2001} in reasonable
accord with our optical properties data. The low-energy dielectric
response, discussed below, is consistent with this picture.

Figure \ref{fig_PrCondReflPrFM}(b) displays a close-up view of the
4.2 K optical conductivity within the $ab$ plane in  zero (SGI) and
high magnetic fields (FMM). Interestingly, the bound carrier
excitation is stabilized in the high field state, moving from
$\sim$40 to 60 meV with a substantial increase in oscillator
strength. To quantify  changes in localization with applied magnetic
field, we fit the spectra with several model oscillators over this
energy range. Both peak position (inset, Fig.
\ref{fig_PrCondReflPrFM}(b)) and spectral weight show a first-order
transition near 4 T, demonstrating that the ferromagnetic domains,
which are associated with the bound carrier localization, couple to
the field-induced transition. We previously conjectured that the
presence of this low-energy bound carrier excitation in \PrLSMO\ may
be connected with the improved magnetoresistance
properties.\cite{Woodward2004} Evidence for domains in the FMM state
also comes from recent resistive relaxation
studies.\cite{Matsukawa2005} Optical imaging techniques provide
direct evidence for texture changes with applied magnetic field as
well.\cite{Tokunaga2005}

\begin{figure}
\includegraphics[width = 3.25 in]{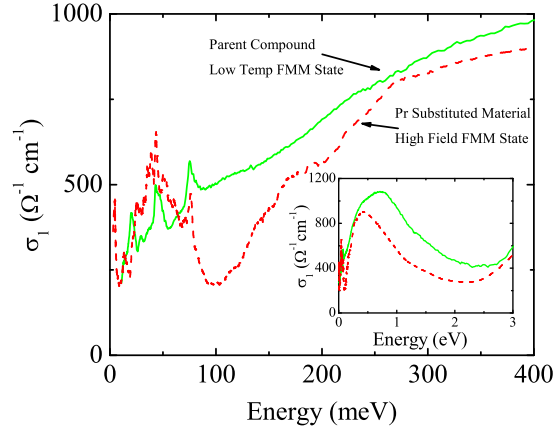}
\caption{\label{fig_PrCompare} (Color online) $ab$ plane optical
conductivity of the low temperature (10 K, 0 T) FMM state of
La$_{1.2}$Sr$_{1.8}$Mn$_2$O$_7$ and the high field (4.2 K, 10T,
$H\parallel c$) FMM state of
(La$_{0.4}$Pr$_{0.6}$)$_{1.2}$Sr$_{1.8}$Mn$_2$O$_7$. The spectrum of
the parent compound is from Ref. \onlinecite{Lee2000}. }
\end{figure}

To gain additional insight into the influence of disorder, it is
useful to compare the low-energy dynamics in the field-induced FMM
state of \PrLSMO\ with the temperature-induced FMM state of the
double-layer parent compound La$_{1.2}$Sr$_{1.8}$Mn$_2$O$_7$. The
latter is  obtained from the literature data of Lee {\it et
al.}.\cite{Lee2000} Neither material is a conventional metal
(Fig.~\ref{fig_PrCompare}). The optical conductivity of
La$_{1.2}$Sr$_{1.8}$Mn$_2$O$_7$ is characteristic of a weak metal,
with a small polaron hopping band (Mn$^{3+}$ to Mn$^{4+}$ charge
transfer) overlapping the onsite Mn $d$ to $d$
transitions.\cite{Lee2000} The high-temperature  $\sim$0.15 - 0.3 eV
pseudogap in the optical conductivity is  attributed to short-range
charge/orbital ordering.\cite{Lee2000} From a theoretical point of
view, pseudogap formation is predicted to be a generic consequence
of mixed-phase regimes.\cite{Moreo1999} This gap  begins to fill
below T$_c$ and disappears  in the low temperature FMM
state.\cite{Lee2000} The response of the Pr-substituted material is
similar, but the far-infrared bound carrier excitation is separated
from the asymmetric set of electronic excitations centered at 0.4 eV
by a smaller pseudogap near 100 meV. The oscillator strength lost
due to pseudogap formation is recovered in the bound carrier peak at
60 meV (Fig.~\ref{fig_PrCompare}).  We conclude that, although the
low temperature state of La$_{1.2}$Sr$_{1.8}$Mn$_2$O$_7$ and the
high field state of the Pr-substituted material are both considered
to be FMMs from the bulk properties point of view, \cite{Gordon2001,
Lee2000} their low-energy electronic response is different. In
ARPES, $k$-space sampling is different and  the pseudogap energy
scale is more difficult to define because the gap is ``soft".   Even
so, in the high temperature phase, the size of the pseudogap is
generally consistent with  optical results in
La$_{1.2}$Sr$_{1.8}$Mn$_2$O$_7$ and related double layer
manganites.\cite{Sun2006, Mannella2005, Lee2000} Interestingly,
ARPES shows that the pseudogap persists into the low-temperature
phase of La$_{1.2}$Sr$_{1.8}$Mn$_2$O$_7$, with an extrapolated value
of $\sim$90 meV.\cite{Chuang2001} This differs from the optical
property results for the parent compound below T$_c$,\cite{Lee2000}
but is in line with the 100 meV pseudogap in
(La$_{0.4}$Pr$_{0.6}$)$_{1.2}$Sr$_{1.8}$Mn$_2$O$_7$
(Fig.~\ref{fig_PrCompare}). Disorder effects in the Pr-substituted
material likely break $k$-space selection rules and change the
optical activity of this excitation.

\begin{figure}[b]
\includegraphics[width = 3.25 in]{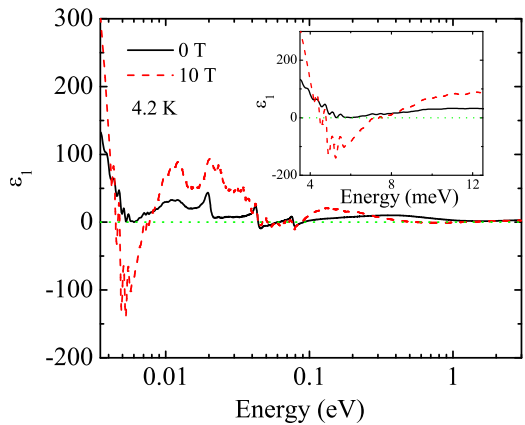}
\caption{\label{fig_PrDielectric} (Color online) $ab$ plane real
part of the dielectric constant of
(La$_{0.4}$Pr$_{0.6}$)$_{1.2}$Sr$_{1.8}$Mn$_2$O$_7$ at 4 K in the
zero (SGI) and high field (FMM, $H\parallel c$) states,   as
determined by Kramers-Kronig analysis.}
\end{figure}

Figure~\ref{fig_PrDielectric} displays the real part of the
dielectric function, emphasizing the strong dielectric contrast
between the low and high field states of
(La$_{0.4}$Pr$_{0.6}$)$_{1.2}$Sr$_{1.8}$Mn$_2$O$_7$.\cite{dielectric_note}
In the SGI state, the dispersive response is typical of a
dielectric, with an overall  positive $\epsilon_1$ except for the
strongest phonon dispersions. The dispersive response of the high
field FMM state of the Pr-substituted compound is completely
different.  It first crosses zero at $\sim$1.2 eV,  a plasma energy
that is typical for a metal. However, it deviates from a Drude-like
response and recrosses into positive territory at $\sim$0.6 eV.
Below 0.01 eV, $\epsilon_1$ turns sharply negative, indicative of
weak metallic behavior, only to climb steeply below 4 meV.  Multiple
crossings of the dielectric function are reminiscent of metallic
polymers, where competition between resonant quantum tunneling among
nanoscale grains in the networks and variable range hopping
processes determine the composite dielectric response.
\cite{Prigodin2003}
The  dielectric behavior of
(La$_{0.4}$Pr$_{0.6}$)$_{1.2}$Sr$_{1.8}$Mn$_2$O$_7$ is very
different from the field-induced dielectric change in
Nd$_{1/2}$Sr$_{1/2}$MnO$_3$ and Pr$_{1/2}$Sr$_{1/2}$MnO$_3$, where
the three-dimensional FMM state is associated with a negative
$\epsilon_1$ in the low energy range. \cite{Jung2000a, Jung2000b}
Based upon the results of Fig.~\ref{fig_PrDielectric}, we anticipate
static magnetodielectric effects on the order of 100\% in
(La$_{0.4}$Pr$_{0.6}$)$_{1.2}$Sr$_{1.8}$Mn$_2$O$_7$, although the
mechanism may be very different from that in  the low-bandwidth
cubic manganite Pr$_{0.7}$Ca$_{0.3}$MnO$_3$, where the enormous
magnetodielectric response at 100 Hz is attributed to a decrease in
polaron activation energy with applied
field.\cite{Freitas2005,janlastnote}

\begin{figure}[h]
\includegraphics[width = 3.5 in]{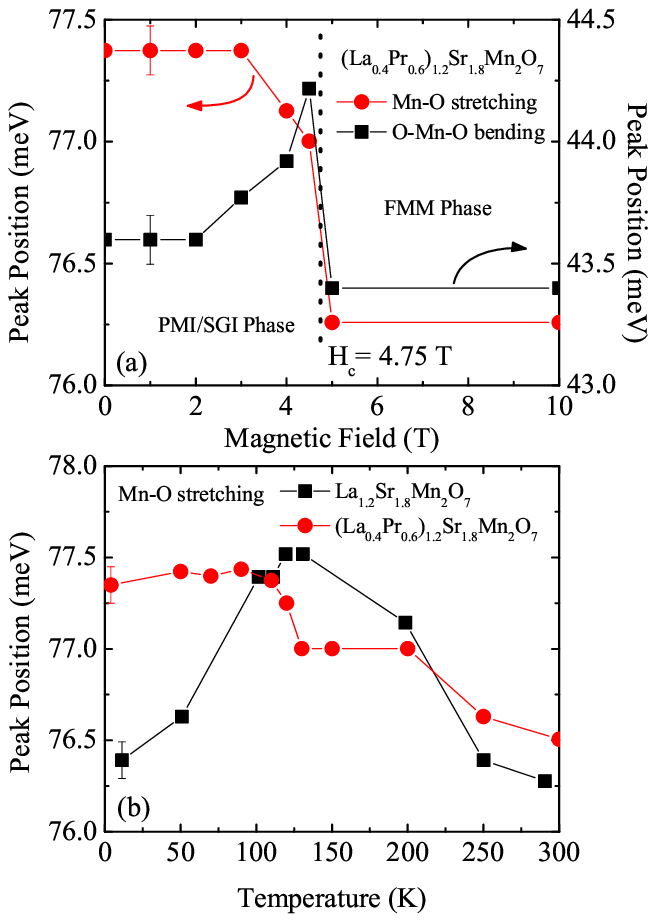}
\caption{\label{fig_PrphononTr} (Color online) (a) Field-dependent
Mn-O stretching and bending modes of
(La$_{0.4}$Pr$_{0.6}$)$_{1.2}$Sr$_{1.8}$Mn$_2$O$_7$. (b)
Temperature-dependence of the Mn-O stretching modes of \PrLSMO\ and
La$_{1.2}$Sr$_{1.8}$Mn$_2$O$_{7}$, the double-layer parent compound
(T$_c$=120 K). Data for La$_{1.2}$Sr$_{1.8}$Mn$_2$O$_{7}$ are from
Lee {\it et al.} in  Ref. \onlinecite{Lee2000}. }
\end{figure}

Direct information on how phonons couple with the magnetically
driven transition in \PrLSMO\ is also of interest, especially since
stretching modes have been implicated in electronic kink formation
in recent photoemission studies of a double-layer
manganite,\cite{Sun2006} and correlation between the Mn-O stretching
mode and polaron formation was demonstrated by optics in
La$_{1.2}$Sr$_{1.8}$Mn$_2$O$_7$.\cite{Lee2000} In addition to the
well-known infrared-active modes of the double-layer
compounds,\cite{Lee2000} the vibrational spectrum of \PrLSMO\
exhibits a number of new, small modes in the low-field insulating
state that are likely derived from the symmetry breaking effects of
Pr substitution.\cite{Woodward2004,Romero1998} Some of these smaller
structures become quite prominent in the FMM state, riding on top of
the far-infrared bound carrier excitation (Fig.
\ref{fig_PrCondReflPrFM}(b)). Several modes shift with applied
field, indicative of a significant structural change between the
low-field insulating to high-field metallic phases.
Figure~\ref{fig_PrphononTr}(a) displays trends in the
field-dependent Mn-O stretching and bending modes. Upon increasing
field, the 77.4 meV Mn-O(3) stretching mode softens on approaching
to the transition, shows a sharp change through H$_c$, and saturates
above 5 T. The 46.3 meV O-Mn-O bending mode is more complicated. It
hardens on approaching the transition and then drops suddenly across
H$_c$. Several other modes soften through the SGI - FMM transition
as well, indicative of a structural change. For comparison,
Fig.~\ref{fig_PrphononTr}(b) displays the temperature dependent
trends of the Mn-O stretching mode in both \PrLSMO\ and
La$_{1.2}$Sr$_{1.8}$Mn$_2$O$_{7}$.\cite{Lee2000} Note that the Mn-O
stretching mode softens through T$_c$ in the parent
compound,\cite{Lee2000} whereas it softens through H$_c$ in the
Pr-substituted material.
We can also compare field-induced modifications in the Mn-O
stretching and bending modes with  neutron diffraction and
magnetostriction results.\cite{Nakanishi, Gukasov2005} Neutron
diffraction data shows  that several bond lengths change with field.
For instance, the distance between Mn and O(3) is 1.931(8) \AA ~in
the SGI state and increases to 1.937(2) \AA ~in the FMM
state.\cite{Wang2003, Gukasov2005} This trend is consistent with our
observation of softer phonon modes in the high field FMM state of
\PrLSMO\ (Fig.~\ref{fig_PrphononTr}(a)). On the other hand,
magnetostriction is a bulk technique. The elastic constants of
\PrLSMO\ soften through the transition and then tighten with applied
field as Mn-O bond length disorder is suppressed.\cite{Nakanishi}

\begin{table}
\caption{\label{tab_radii}} The averaged radii $r_A$ and variance of
the ionic radii $\sigma^2$ of
(La$_{1-z}$Pr$_{z}$)$_{1.2}$Sr$_{1.8}$Mn$_2$O$_7$, calculated  at
the perovskite  (P) and rock salt (R) sites.
\begin{ruledtabular}

\begin{tabular}{lcllll}
 &&\multicolumn{2}{c}{P Sites}&\multicolumn{2}{c}{R Sites}\\

    \hline
$z$ & T$_c$ (K)  & $r_A$ (\AA)& $\sigma^2$ (\AA$^2$)& $r_A$ (\AA)& $\sigma^2$ (\AA$^2$)\\
\hline

0   & 120 &1.408&0.00154&1.272&0.00212 \\
0.2 & 90  &1.403&0.00226&1.269&0.00256 \\
0.4 & 60  &1.398&0.00294&1.266&0.00297 \\
0.6 &   - &1.394&0.00358&1.264&0.00337 \\
\end{tabular}
\end{ruledtabular}
\footnotemark{The averaged radii $r_A$ and the variance of the radii
$\sigma^2$ are calculated as $r_A=\sum x_ir_i$ and $\sigma^2=\sum
x_i(r_i^2-r_A^2)$, where $x_i$ and $r_i$ are the fractional
occupancies ($\sum x_i=1$) and effective ionic radii of rare earth
and alkaline earth cations, respectively.
\cite{Tomioka2004,Rodiguez1996}}

\footnotemark{Standard ionic radii are obtained from Refs.
\onlinecite{Pollert1982} and \onlinecite{Shannon1976}. The P site
has a coordination number of 12, and the R site has a coordination
number of 9. }
\end{table}

Tomioka and Tokura recently proposed a global phase diagram of
perovskite manganites within  the plane of
 effective one-electron
bandwidth and  quenched disorder.\cite{Tomioka2004} These two
parameters are controlled by the average value and the variance of
rare earth and alkaline earth ionic radii, respectively. With
increasing quenched disorder, the Y-shaped bicritical feature of the
phase diagram splits and a SGI state emerges, separating the FMM and
charge/orbital ordered antiferromagnetic phases. The authors propose
that spin-glass character may enhance the colossal magnetoresistance
effect in these materials.\cite{Tomioka2004} Reviewing complexity in
strongly correlated electronic systems, Dagotto emphasizes that this
glassy region is dominated by short-range correlations and a local
tendency towards either FMM or AFI regimes.\cite{Burgy2001,
Dagotto2005}

Although the Tomioka-Tokura phase diagram was developed for cubic
perovskites,\cite{Tomioka2004} the underlying physics may be useful
for understanding the double-layer manganites as well. We therefore
investigated trends in
(La$_{1-z}$Pr$_{z}$)$_{1.2}$Sr$_{1.8}$Mn$_2$O$_7$ within the general
framework of this global electronic phase diagram picture, modified
to account for the two distinct perovskite (P) and rock salt (R)
sites in the double-layer material. Table~\ref{tab_radii} shows
calculated values of the averaged radii ($r_A$) and the variance of
the ionic radii ($\sigma^2$)  with different Pr substitution.  Both
P and R sites show the same trend:
 $r_{A}$ decreases and $\sigma^2$
increases with  Pr substitution. Can we understand bulk property
trends in \PrLSMO\ within this picture? Table~\ref{tab_radii} shows
the correlation between the PMI to FMM transition temperature
(T$_c$),\cite{Apostu2001, Matsukawa2003} electronic bandwidth, and
quenched disorder. With increasing Pr substitution, T$_c$ decreases
and is eventually quenched.\cite{Apostu2001, Matsukawa2003} Despite
the suppression of T$_c$ at $z$=0.6, \PrLSMO\ is still in close
proximity to the FMM phase boundary. Thus, a small local structure
variation can  modify $r_A$ and $\sigma^2$, pushing the system back
into the FMM state. The measured phonon softening through the H$_c$
(Fig.~\ref{fig_PrphononTr}(a)) demonstrates that local Mn-O bonding
is more relaxed in the high field FMM state, in line with
observations by Lee {\it et al.} of a relaxed lattice in the
double-layer parent compound at low temperature.\cite{Lee2000} It is
therefore probable that the applied field modifies $r_A$ and
overcomes substitution-induced disorder effects in $\sigma^2$,
driving the system back into the FMM state in the Tomioka-Tokura
phase diagram. The principal difference between the field-induced
FMM state in \PrLSMO\ and the temperature-induced FMM state in
La$_{1.2}$Sr$_{1.8}$Mn$_2$O$_{7}$ is the presence of the pseudogap
in the substituted material (Fig. \ref{fig_PrCompare}). The
anticipated curvature of the SGI/FMM phase boundary in magnetic
field is consistent with this picture and facilitates the
magnetically-driven transition. At the same time, the low energy
scale of H$_c$ indicates the closely competing nature of the SGI and
FMM phases in the Pr-substituted double-layer manganite. These
results suggest that control of disorder in the proximity of a phase
boundary may provide an important route to magnetodielectric
materials that are switchable in low magnetic fields.

\subsection{Optical Phase Diagram}

\begin{figure}
\includegraphics[width = 3.5 in]{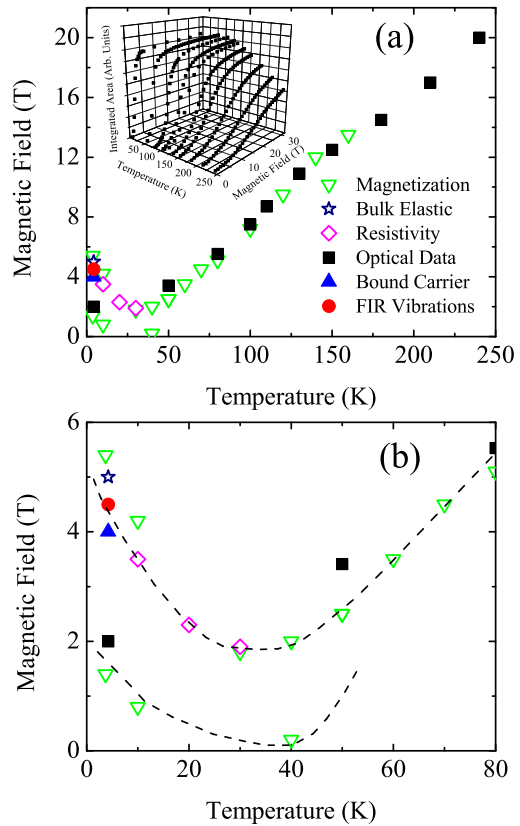}
\caption{\label{fig_PrHTPhTr} (Color online) (a) H-T phase diagram,
extracted from the optical properties data for H $\parallel$ $c$.
Data are taken with increasing field. Selected magnetization,
magnetostriction, and resistivity results are shown for
comparison.\cite{Gordon2001,Matsukawa2004,Nakanishi} The inset
displays the absolute value of the integrated area of the
reflectance ratio spectra in the color band region as a function of
temperature and magnetic field. Inflection points in the data were
used to determine phase boundary locations. (b) Close-up of view of
the H-T phase diagram in the low temperature region.}
\end{figure}

Our comprehensive magneto-optical measurements allow us to generate
the H-T phase diagram of \PrLSMO\ for $H\parallel c$
(Fig.~\ref{fig_PrHTPhTr}).\cite{extractphasediagram} For comparison,
we also plotted several points determined from selected resistivity,
magnetization, and magnetostriction measurements.
\cite{Gordon2001,Matsukawa2004,Nakanishi} Magnetization measurements
indicate two boundaries at $\sim$2 and 5 T below 50 K, perhaps due
to domain rotation.\cite{Nakanishi} Above 50 K, relaxation effects
are more rapid.\cite{Gordon2001} Here,  the phase diagram from both
optics (which measures the microscopic nature of the charge degrees
of freedom) and magnetization are in good agreement, although the
boundary becomes more diffuse with increasing temperature (inset,
Fig.~\ref{fig_PrHTPhTr}(a)).

It is especially interesting to compare the correspondence of the
low temperature phase boundaries in
(La$_{0.4}$Pr$_{0.6}$)$_{1.2}$Sr$_{1.8}$Mn$_2$O$_7$, determined by
different techniques. Recently, Matsukawa {\it et al.} used combined
resistivity, magnetization, and magnetostriction measurements to
show that the bulk relaxation time of the lattice is two orders of
magnitude smaller than that of the resistivity.\cite{Matsukawa2005}
Extending this analysis to include microscopic trends, we find that
the critical field extracted from field-induced changes in Mn$^{3+}$
$e_g$ orbital population is similar to the first magnetization
boundary ($\sim$2 T). On the other hand, direct measurements of Mn-O
and O-Mn-O stretching and bending vibrational modes
(Fig.~\ref{fig_PrphononTr}(a)) show excellent correspondence with
critical fields determined from bulk magnetostriction, resistivity,
and  the second magnetization boundary ($\sim$5 T). Changes in the
far-infrared bound carrier excitation (inset,
(Fig.~\ref{fig_PrCondReflPrFM}(b)) are also associated with this
lattice distortion. The results demonstrate that the lattice
responds more slowly than charge and spin over various length and
time scales. The strong hysteresis in the low temperature physical
properties\cite{Apostu2001,Gordon2001,Choi2004a, Matsukawa2004,
Matsukawa2005, Nakanishi} of \PrLSMO\ derives from these
differences.

%
%

\subsection{Field-Induced Color Changes at 300 K and Visualization}

\begin{figure}[b]
\includegraphics[width = 3.5 in]{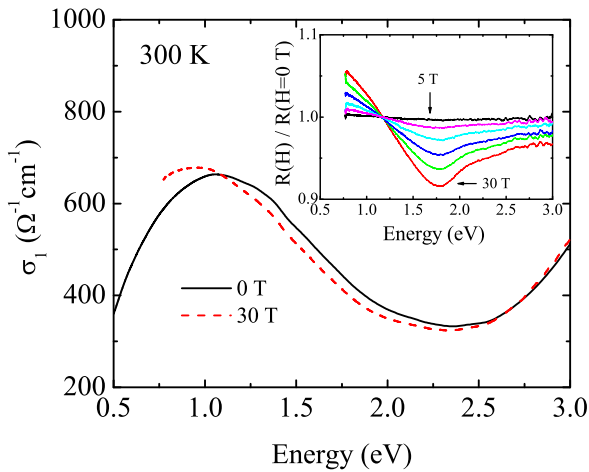}
\caption{\label{fig_PrCondRT} (Color online) 300 K $ab$ plane
optical conductivity of \PrLSMO\ in 0 (solid line) and 30 T (dashed
line) magnetic fields ($H\parallel c$). The inset shows the measured
$ab$ plane reflectance ratio at 5, 10, 15, 20, 25, and 30 T (top to
bottom). The data are taken upon increasing field.}
\end{figure}

Controlled disorder near a phase boundary can clearly yield a
substantial magnetodielectric response over a wide energy range.
However, at this time,  most magnetodielectric materials
\cite{Cruz2005, Lorenz2004, Goto2004, Rogado2005, Choi2004a,
Choi2004b, Woodward2005, Rai2006,Jung2000a, Jung2000b,Freitas2005}
display  contrast only at low temperatures. Strategic control of
spin-lattice interactions or the presence of a magnetically-driven
phase transition (and associated change in ground state) offer
potential routes to higher transition temperatures. The latter is
investigated here. Specifically, we explored whether the  remnants
of the SGI-FMM transition in \PrLSMO\
 can be used to drive color property changes at room temperature.

Figure~\ref{fig_PrCondRT} displays the room temperature optical
conductivity of \PrLSMO\ at 0 and 30 T. The field-induced spectral
changes are similar to but smaller than those at base temperature
(Fig.~\ref{fig_PrCondReflPrFM}). Under applied magnetic field, the
entire visible spectrum redshifts. Given high enough magnetic field,
the full SGI-FMM transition can likely be realized, inducing
magneto-dielectric effects similar to those at 4 K. Recent
$^{139}$La NMR studies reveal  field-induced long-range
ferromagnetic order up to 330 K in
La$_{1.2}$Sr$_{1.8}$Mn$_2$O$_{7}$,\cite{Shiotani2006} suggesting
that 300 K magnetochromic effects may also be present in the parent
compound.

\begin{figure} [width= 4 in]
\includegraphics{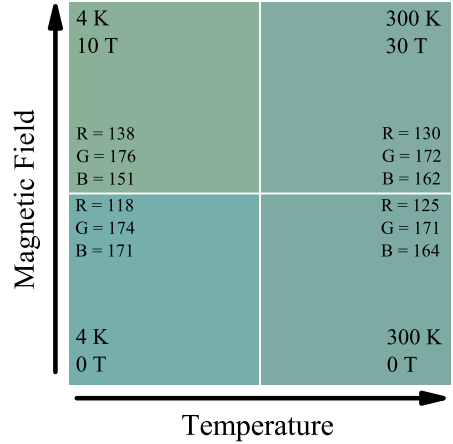}
\caption{\label{fig_PrColor}(Color online) Schematic view of the
temperature and magnetic field-induced color changes in
(La$_{0.4}$Pr$_{0.6}$)$_{1.2}$Sr$_{1.8}$Mn$_2$O$_7$. A comparison to
transmittance data allows an estimated scaling constant of
K=1.22x10$^{-5}$ cm to be determined. This K value was used in the
calculation of the other colors. The RGB values for each swatch are
given as well.}
\end{figure}

We employed standard color rendering techniques to visualize the
magnetochromic and thermochromic effects in \PrLSMO\
(Fig.~\ref{fig_PrColor}).\cite{Billmeyer2000,Color_Parameters}
Examination of the color panel shows a distinct color change between
the  0 and 10 T data at low temperature, whereas there is only
modest thermochromism as temperature is increased. The 300 K data
taken at 30 T show a small change in color compared to that of the 0
T room temperature data. This shows that the color change is
quenched by temperature although larger fields can probably drive
the SGI to FMM transition at 300 K. In this case, we anticipate
magnetochromic effects similar to those at low temperature. The
calculated RGB values are presented (Fig.~\ref{fig_PrColor}) to
quantify these color properties.

\section{Conclusion}

We report a comprehensive magneto-optical investigation of \PrLSMO,
undertaken to examine the microscopic aspects of spin, lattice,
charge, and orbital coupling through the SGI to FMM transition.
Overall, spectral weight shifts to lower energy with applied
magnetic field, although the material displays far-infrared
localization and a pseudogap due to ferromagnetic domain formation
and modest dc conductivity in the high field FMM state, different
from that in the double-layer parent material. Several Mn-O and
O-Mn-O stretching and bending vibrational modes soften through the
field-driven transition, indicative of a structural change.   Using
the Tomioka-Tokura phase diagram picture\cite{Tomioka2004}, we find
that the FMM state, although quenched by chemical doping, is revived
in a modified form by the application of magnetic field by driving
the system from the disordered SGI phase into the FMM regime. The
optical properties were also used to map the H-T phase diagram, and
we find that the lattice response to the field is slower than that
of spin and charge. Finally, we demonstrate that remnants of the
SGI-FMM transition can drive 300 K color changes  in
(La$_{0.4}$Pr$_{0.6}$)$_{1.2}$Sr$_{1.8}$Mn$_2$O$_7$.  Color
rendering provides a visual representation of the spectral changes.
It is of future interest to complement this work with static
magnetodielectric studies.

\section{ACKNOWLEDGMENTS}

Work at the University of Tennessee is supported by the Materials
Science Division, Basic Energy Sciences, U.S. Department of Energy
(DE-FG02-01ER45885). The international aspects of this research were
supported by the National Science Foundation (INT-019650). A portion
of this work was performed at the NHMFL, which is supported by NSF
Cooperation Agreement DMR-0084173 and by the State of Florida. We
thank E. Dagotto (UT/ORNL) and D. Dessau (Colorado) for useful
discussions.


\begin{thebibliography}{JinboCao}


\bibitem{Tokura2000}
{\it Colossal Magnetoresistive Oxides}, edited by Y. Tokura (Gordon
and Breach Science Publishers, New York, 2000).


\bibitem{Dagotto2003}
E. Dagotto, {\it Nanoscale Phase Separation and Colossal
Magneto-resistance}, (Springer, 2003).


\bibitem{Chatterji2004}
{\it Colossal Magnetoresistive Manganites}, edited by T. Chatterji
(Kluwer Academic Publishers, 2004)


\bibitem{Kubota2000}
M. Kubota, H. Fujioka, K. Hirota, K. Ohoyama, Y. Moritomo, H.
Yoshizawa, and Y. Endoh, J. Phys. Soc. Jpn. \textbf{69}, 1606
(2000).


\bibitem{Apostu2001}
M. Apostu, R. Suryanarayanan, A. Revcolevschi, H. Ogasawara, M.
Matsukawa, M. Yoshizawa, and N. Kobayashi, Phys. Rev. B \textbf{64},
012407 (2001).


\bibitem{Moritomo1996}
Y. Moritomo, A. Asamitsu, H. Kuwahara, and Y. Tokura, Nature
(London) \textbf{380}, 141 (1996).


\bibitem{Mitchell1997}
J. F. Mitchell, D. N. Argyriou, J. D. Jorgensen, D. G. Hinks, C. D.
Potter, and S. D. Bader, Phys. Rev. B \textbf{55}, 63 (1997).


\bibitem{Perring1997}
T. G. Perring, G. Aeppli, Y. Moritomo, and Y. Tokura, Phys. Rev.
Lett. \textbf{78}, 3197 (1997).


\bibitem{Cruz2005}
C. dela Cruz, F. Yen, B. Lorenz, Y. Q. Wang, Y. Y. Sun, M. M.
Gospodinov, and C. W. Chu, Phys. Rev. B \textbf{71}, R060407 (2005).


\bibitem{Lorenz2004}
B. Lorenz, A. P. Litvinchuk, M. M. Gospodinov, and C. W. Chu, Phys.
Rev. Lett. \textbf{92}, 087204 (2004).


\bibitem{Goto2004}
T. Goto, T. Kimura, G. Lawes, A. P. Ramirez, and Y. Tokura, Phys.
Rev. Lett. \textbf{92}, 257201 (2004).


\bibitem{Rogado2005}
N. S. Rogado, J. Li, A. W. Sleight, and M. A. Subramanian, Adv.
Mater. \textbf{17}, 2225 (2005).


\bibitem{Choi2004a}
J. Choi, J. D. Woodward, J. L. Musfeldt, J. T. Haraldsen, X. Wei, M.
Apostu, R. Suryanarayanan, and A. Revcolevschi, Phys. Rev. B
\textbf{70}, 064425 (2004).


\bibitem{Choi2004b}
J. Choi, J. D. Woodward, J. L. Musfeldt, X. Wei, M. H. Whangbo, J.
He, R. Jin, and D. Mandrus, Phys. Rev. B \textbf{70}, 085107 (2004).


\bibitem{Woodward2005}
J. D. Woodward, J. Choi, J. L. Musfeldt, J. T. Haraldsen, X. Wei, H.
-J. Koo, D. Dai, M. -H. Whangbo, C. P. Landee, and M. M. Turnbull,
Phys. Rev. B \textbf{71}, 174416 (2005).


\bibitem{Rai2006}
R. C. Rai, J. Cao, J. L. Musfeldt, D. J. Singh, X. Wei, R. Jin, Z.
X. Zhou, B. C. Sales, and D. Mandrus, Phys. Rev. B \textbf{73},
075112 (2006).


\bibitem{Lee2002}
H. J. Lee, K. H. Kim, M. W. Kim, T. W. Noh, B. G. Kim, T. Y. Koo,
S.-W. Cheong, Y. J. Wang, and X. Wei, Phys. Rev. B \textbf{65},
115118 (2002).


\bibitem{Alexandrov2000}
A. S. Alexandrov and A. M. Bratkovsky, J. Appl. Phys. \textbf{87},
5016 (2000).


\bibitem{Jung2000a}
J. H. Jung, H. J. Lee, T. W. Noh, E. J. Choi, Y. Moritomo, Y. J.
Wang, and X. Wei, Phys. Rev. B \textbf{62}, 481 (2000).


\bibitem{Jung2000b}
J. H. Jung, H. J. Lee, T. W. Noh, Y. Moritomo, Y. J. Wang, and X.
Wei, Phys. Rev. B \textbf{62}, 8634 (2000).


\bibitem{Okimoto1998}
Y. Okimoto, Y. Tomioka, Y. Onose, Y. Otsuka, and Y. Tokura, Phys.
Rev. B \textbf{57}, R9377 (1998).


\bibitem{Okimoto1999}
Y. Okimoto, Y. Tomioka, Y. Onose, Y. Otsuka, and Y. Tokura, Phys.
Rev. B \textbf{59}, 7401 (1999).


\bibitem{Freitas2005}
R. S. Freitas, J. F. Mitchell, and P. Schiffer, Phys. Rev. B
\textbf{72}, 144429 (2005).


\bibitem{Tomioka2004} Y. Tomioka, and Y. Tokura, Phys. Rev. B
\textbf{70}, 014432 (2004).


\bibitem{Alexandrov1999} A. S. Alexandrov and A. M. Bratkovsky, Phys.
Rev. B. \textbf{60}, 6215 (1999).


\bibitem{Matsukawa2003}
M. Matsukawa, M. Narita, T. Nishimura, M. Yoshizawa, M. Apostu, R.
Suryanarayanan, A. Revcolevschi, K. Itoh, and N. Kobayashi, Phys.
Rev. B \textbf{67}, 104433 (2003).


\bibitem{Wang2003}
F. Wang, A. Gukasov, F. Moussa, M. Hennion, M. Apostu, R.
Suryanarayanan, and A. Revcolevschi, Phys. Rev. Lett. \textbf{91},
047204 (2003).


\bibitem{Woodward2004} J. D. Woodward, J. Choi, J. L. Musfeldt, J. T.
Haraldsen, M. Apostu, R. Suryanarayanan, and A. Revcolevschi, Phys.
Rev. B \textbf{69}, 104415 (2004).


\bibitem{Gordon2001}
I. Gordon, P. Wagner, V. V. Moshchalkov, Y. Bruynseraede, M. Apostu,
R. Suryanarayanan, and A. Revcolevschi, Phys. Rev. B \textbf{64},
092408 (2001).


\bibitem{Wagner2002}
P. Wagner, I. Gordon, V. V. Moshchalkov, Y. Bruynseraede, M. Apostu,
R. Suryanarayanan, and A. Revcolevschi, Europhys. Lett \textbf{58},
285 (2002).


\bibitem{Matsukawa2004} M. Matsukawa, M. Chiba, A. Akasaka, R. Suryanarayanan,
M. Apostu, A. Revcolevschi, S. Nimori, and N. Kobayashi, Phys. Rev.
B \textbf{70}, 132402 (2004).


\bibitem{Matsukawa2005} M. Matsukawa, K. Akasaka, H. Noto, R. Suryanarayanan,
 S. Nimori, M. Apostu, A. Revcolevschi, and N. Kobayashi, Phys. Rev. B
\textbf{72}, 064412 (2005).


\bibitem{Moussa2004}
F. Moussa, M. Hennion, F. Wang, A. Gukasov, R. Suryanarayanan, M.
Apostu, and A. Revcolevschi, Phys. Rev. Lett. \textbf{93}, 107202
(2004).


\bibitem{Gukasov2005} A. Gukasov, F. Wang, B. Anighoefer, L. He,
R. Suryanarayanan, and A. Revcolevschi, Phys. Rev. B \textbf{72},
092402 (2005).


\bibitem{Tokunaga2005} Y. Tokunaga, M. Tokunaga, and T. Tamegai, Phys. Rev. B
\textbf{71}, 012408 (2005).


\bibitem{Wooten1972}
F. Wooten, {\it Optical Properties of Solids} (Academic Press, New
York, 1972).


\bibitem{Billmeyer2000} F. W. Billmeyer and M. Saltzman, \textit{Principles of Color Technology,
3rd Ed.}, Wiley, New York (2000).


\bibitem{Musfeldt1993} J. L. Musfeldt, D. B. Tanner, and A. J. Paine, J. Opt. Soc. Am. A \textbf{10}
2648 (1993).


\bibitem{absorption} Through a Kramers-Kronig analysis, one can determine the
extinction coefficient as a function of frequency, $\kappa(\omega)$.
The absorption coefficient, $\alpha$, is calculated as
$\alpha$=$4\pi \kappa(\omega)$.


\bibitem{K}The absorption coefficient needs to be
normalized by a constant, K, to determine the effective absorption
of the material. K is dependent on factors such as the mass fraction
and thickness of the transmittance sample. However, because these
are reflectance measurements of a solid material, K is unknown. This
constant can be approximated by an examination of the transmittance
of the material or by normalizing the absorption to a distinct value
of color (assuming the color of the material is known). Once K is
determined for a material, it is the same for all spectra and is not
dependent on magnetic field or temperature.


\bibitem{Ishikawa1998}
T. Ishikawa, T. Kimura, T. Katsufuji, and Y. Tokura, Phys. Rev. B
\textbf{57}, R8079 (1998).


\bibitem{Lee2000}
H. J. Lee, K. H. Kim, J. H. Jung, T. W. Noh, R. Suryanarayanan, G.
Dhalenne, and A. Revcolevschi, Phys. Rev. B \textbf{62}, 11320
(2000).


\bibitem{Moreo1999}
A. Moreo, S. Yunoki, and E. Dagotto, Phys. Rev. Lett. \textbf{83},
2773 (1999).


\bibitem{Sun2006}
Z. Sun, Y. -D Chuang, A. V. Fedorov, J. F. Douglas, D. Reznik, F.
Weber, N. Aliouane, D. N. Argyriou, H. Zheng, J. F. Mitchell, T.
Kimura, Y. Tokura, A. Revcolevschi, and D. S. Dessau,
arXiv:cond-mat/0510255 v1, 10 Oct 2005.


\bibitem{Mannella2005}
N. Mannella, W. L. Yang, X. J. Zhou, H. Zheng, J. F. Mitchell, J.
Zaanen, T. P. Devereaux, N. Nagaosa, Z. Hussain, and Z.-X. Shen,
Nature \textbf{438}, 474 (2005).


\bibitem{Chuang2001}
Y.-D. Chuang, A. D. Gromko, D. S. Dessau, T. Kimura, and Y. Tokura,
Science \textbf{292}, 1509 (2001).


\bibitem{dielectric_note}
In contrast, $\epsilon_1$ of the parent compound in the FMM state is
positive within the whole frequency range, increasing upon
approaching to zero energy, following a typical ``bad metal''
behavior.\cite{Lee2000}


\bibitem{Prigodin2003}
V. N. Prigodin and A. J. Epstein, Physica B: Condensed Matter
\textbf{338}, 310 (2003).


\bibitem{janlastnote}
This is because the polaron excitation remains well-defined in the
high field FMM state of
(La$_{0.4}$Pr$_{0.6}$)$_{1.2}$Sr$_{1.8}$Mn$_2$O$_7$
(Fig.~\ref{fig_PrCondReflPrFM}(a)).


\bibitem{Romero1998}
D. B. Romero, V. B. Podobedov, A. Weber, J. P. Rice, J. F. Mitchell,
R. P. Sharma, and H. D. Drew, Phys. Rev. B \textbf{58}, R14737
(1998).


\bibitem{Nakanishi}
Y. Nakanishi, K. Shimomura, T. Kumagai, M. Matsukawa, M. Yoshizawa,
M. Apostu, R. Suryanarayanan, A. Revcolevcschi, and S. Nakamura,
(unpublished).


\bibitem{Rodiguez1996}
L. M. Rodriguez-Martinez and J. P. Attfield, Phys. Rev. B
\textbf{54}, R15622 (1996).


\bibitem{Burgy2001}
J. Burgy, M. Mayr, V. Martin-Mayor, A. Moreo, and E. Dagotto, Phys.
Rev. Lett. \textbf{87}, 277202 (2001).


\bibitem{Dagotto2005}
E. Dagotto, Science \textbf{309}, 257 (2005).




\bibitem{Pollert1982}
E. Pollert, S. Krupi\v cka, and E. Kuzmi\v cov\'a, J. Phys. Chem.
Solids \textbf{43}, 1137 (1982).


\bibitem{Shannon1976}
R. D. Shannon, Acta Crystallogr. Sec. A \textbf{32}, 751 (1976).


\bibitem{extractphasediagram}
We measured the magneto-optical response of \PrLSMO\ as a function
of temperature and field, and quantified the changes by plotting the
absolute value of the integrated intensities of the field-induced
feature. The first derivative of this change determined the boundary
from the optical perspective.


\bibitem{Shiotani2006}
Y. Shiotani, J. L. Sarrao, and Guo-qing Zheng, Phys. Rev. Lett.
\textbf{96}, 057203 (2006).




\bibitem{Color_Parameters} In the calculation of color, the
300 K (0 T) data were normalized to K=1.22$\times$10$^{-5}$ cm,
\cite{K} chosen by a comparison to room temperature transmittance.
The same value of K was used to normalize the other data, allowing
us to compare overall color changes to those of the 300 K, zero
field data.



\end{thebibliography}
\end{document}